\documentclass[letterpaper]{JHEP3}
\usepackage{amssymb,amsfonts}
\usepackage{cite}
\usepackage{epsfig}
\newcommand{\trace}{\mbox{Tr}}

\newcommand{\R}{{\bf R}}

\newcommand{\CD}{{\cal D}}
\newcommand{\CF}{{\cal F}}
\newcommand{\CG}{{\cal G}}
\newcommand{\CH}{{\cal H}}

\newcommand{\CL}{{\cal L}}
\newcommand{\CM}{{\cal M}}
\newcommand{\CO}{{\cal O}}

\newcommand{\bk}{{\bf k}}
\newcommand{\bx}{{\bf x}}
\newcommand{\by}{{\bf y}}

\newcommand{\p}{\partial}

\renewcommand{\tilde}[1]{\widetilde{#1}}
\newcommand{\be}{\begin{equation}}
\newcommand{\ee}{\end{equation}}
\newcommand{\bea}{\begin{eqnarray}}
\newcommand{\eea}{\end{eqnarray}}
\newcommand{\ie}{{\it i.e.}}
\newcommand{\eg}{{\it e.g.}}
\usepackage{graphicx}
\usepackage{latexsym}
\newcommand{\dif}{{\rm Diff}_{\!\CF}^{}}
\newcommand{\lambdaw}{\Lambda_W^{}}
\newcommand{\newton}{G_{\!N}^{}}

\title{Quantum Gravity at a Lifshitz Point}
\author{Petr Ho\v{r}ava\\
Berkeley Center for Theoretical Physics and Department of Physics\\
University of California, Berkeley, CA, 94720-7300\\
and\\
Theoretical Physics Group, Lawrence Berkeley National Laboratory\\
Berkeley, CA 94720-8162, USA}
\abstract{We present a candidate quantum field theory of gravity with 
dynamical critical exponent equal to $z=3$ in the UV\@.  (As in condensed 
matter systems, $z$ measures the degree of anisotropy between space and 
time.)  This theory, which at short distances describes interacting 
nonrelativistic gravitons, is power-counting renormalizable in $3+1$ 
dimensions.  When restricted to satisfy the condition of detailed balance, 
this theory is intimately related to topologically massive gravity in three 
dimensions, and the geometry of the Cotton tensor.  At long distances, this 
theory flows naturally to the relativistic value $z=1$, and could therefore 
serve as a possible candidate for a UV completion of Einstein's general 
relativity or an infrared modification thereof.  The effective speed of light, 
the Newton constant and the cosmological constant all emerge from relevant 
deformations of the deeply nonrelativistic $z=3$ theory at short distances.}  
\begin{document}
\section{Introduction}

In recent decades, string theory has become the dominant paradigm for 
addressing questions of quantum gravity.  There are many indications 
suggesting that string theory is sufficiently rich to contain the answers 
to many puzzles, such as the information paradox or the statistical 
interpretation of black hole entropy.  Yet, string theory is also a rather 
large theory, possibly with a huge landscape of vacua, each of which 
leads to a scenario for the history of the universe which may or may not 
resemble ours.  Given this richness of string theory, it might even be 
logical to adopt the perspective in which string theory is not a candidate 
for a unique theory of the universe, but represents instead a natural 
extension and logical completion of quantum field theory.  In this picture, 
string theory would be viewed -- just as quantum field theory -- as a powerful 
technological framework, and not as a single theory.  

If string theory is such an apparently vast structure, it seems natural to ask 
whether quantum gravitational phenomena in $3+1$ spacetime dimensions can be 
studied in a self-contained manner in a ``smaller'' framework.  A useful 
example of such a phenomenon is given by Yang-Mills gauge theories in $3+1$ 
dimensions.  While string theory is clearly a powerful technique for studying 
properties of Yang-Mills theories, their embedding into string theory 
is not required for their completeness: In $3+1$ dimensions, they are UV 
complete in the framework of quantum field theory.  

In analogy with Yang-Mills, we are motivated to look for a ``small'' theory 
of quantum gravity in $3+1$ dimensions, decoupled from strings.  One attempt 
to construct such a ``small'' theory is offered by loop quantum gravity.  In 
this paper, we present a new strategy for addressing this problem.  Compared 
to the previous approaches to quantum gravity, the novelty of our approach is 
that it takes advantage of theoretical concepts developed in recent decades in 
condensed matter physics, in particular in the theory of quantum critical 
phenomena.  

In the context of quantum field theory, the main obstacle against 
perturbative renormalizability of Einstein's theory of gravity in $3+1$ 
dimensions is well-understood (see, \eg , \cite{weinberg} for an excellent 
introduction).  The main problem is that the gravitational coupling constant 
$\newton$ is dimensionful, with a negative dimension $[\newton]=-2$ in mass 
units.   The Feynman rules also involve the graviton propagator, which scales 
with the four-momentum $k_\mu\equiv(\omega,\bk)$ schematically as
\be
\frac{1}{k^2},  
\ee
where $k=\sqrt{\omega^2-\bk^2}$.  At increasing loop orders, the Feynman 
diagrams of this theory require counterterms of ever-increasing degree in 
curvature.  The resulting theory can still be treated as an effective field 
theory, but it requires a UV completion.  Usually, this completion is assumed 
to take the form of string or M-theory.    

An improved UV behavior can be obtained if relativistic higher-derivative 
corrections are added to the Lagrangian (see \cite{ftreport} for a review of 
higher-derivative gravity).  Terms quadratic in curvature not only yield new 
interactions (with a dimensionless coupling), they also modify the 
propagator.  Schematically, we get
\be
\label{resprop}
\frac{1}{k^2}+\frac{1}{k^2}\,\newton k^4\,\frac{1}{k^2}+
\frac{1}{k^2}\,\newton k^4\,\frac{1}{k^2}\,\newton k^4\,\frac{1}{k^2}+\ldots=
\frac{1}{k^2-\newton k^4}
\ee
At high energies, the propagator is dominated by the $1/k^4$ term.  This cures 
the problem of UV divergences, and in fact the calculations in Euclidean 
signature suggest that the theory exhibits asymptotic freedom.  However, this  
cure simultaneously produces a new pathology, which prevents this modified 
theory from being a solution to the problem of quantum gravity: The 
resummed propagator (\ref{resprop}) exhibits two poles,
\be
\frac{1}{k^2-\newton k^4}=\frac{1}{k^2}-\frac{1}{k^2-1/\newton}.  
\ee
One describes candidate massless gravitons, but the other corresponds to 
ghost excitations and implies violations of unitarity, at least in 
perturbation theory.   

Recently, a new class of gravity models was introduced in \cite{mqc}.  These 
models exhibit scaling properties which are anisotropic between space and 
time.  Such an anisotropic scaling is common in condensed matter systems, 
where the degree of anisotropy between space and time is characterized by the 
``dynamical critical exponent'' $z$.  (Relativistic systems automatically 
satisfy $z=1$ as a consequence of Lorentz invariance.)  In models of gravity 
with anisotropic scaling, the problem of renormalizability of gravity is put 
in a new context.  Consider, for example, the case of gravity with $z=2$ 
studied in \cite{mqc}.  As a consequence of the nonrelativistic value of $z$, 
the dimension of the gravitational coupling constant changes.  The critical 
dimension in which the coupling is dimensionless shifts to $2+1$, making the 
system a suitable candidate for describing the worldvolume theory on a bosonic 
membrane.  

The techniques used in the construction of gravity models with anisotropic 
scaling in \cite{mqc} closely parallel methods developed in the 
theory of dynamical critical systems \cite{halperin,ma} and quantum 
criticality \cite{sachdev}.  The prototype of the class of condensed-matter 
models relevant here is the theory of a Lifshitz scalar in $D+1$ dimensions 
\cite{lifshitz,hornreich}, first proposed as a description of tricritical 
phenomena involving spatially modulated phases  (and reviewed in 
\cite{cym,mqc}, see also \cite{ardonne,lubensky}).  The action of the Lifshitz 
scalar is
\be
S=\int dt\,d^D\bx\left\{(\dot\Phi)^2-(\Delta\Phi)^2\right\},
\ee
where ``$\dot{\ \ }$'' denotes the time derivative, and 
$\Delta\equiv\p_i\p_i$ is the spatial Laplacian.  This action describes a 
free-field fixed point with anisotropic scaling and $z=2$.  At this fixed 
point, we can add a relevant deformation to the action,
\be
\label{reld}
-c^2\int dt\,d^D\bx\,\,\p_i\Phi\p_i\Phi.
\ee
Under the influence of this deformation, the theory flows in the infrared to 
$z=1$, with Lorentz invariance emerging as an accidental symmetry at long 
distances.  Note that from the short-distance point of view, the emergent 
long-distance speed of light $c$ originates from the dimensionful coupling 
constant associated with the relevant deformation (\ref{reld}) of the $z=2$ 
fixed point.%
\footnote{The idea that Lorentz invariance might be an emergent symmetry has 
a long history, going back at least to the pioneering paper \cite{nielsen}.}

In our approach to quantum gravity, we consider systems whose scaling 
at short distances exhibits a strong anisotropy between space and time, 
with $z>1$.  This will improve the short-distance behavior of the theory.  
The price we have to pay is that  our theory will not exhibit relativistic 
invariance at short distances.  In fact, many developments in string theory 
suggest that giving up Lorentz invariance as a fundamental symmetry may not be 
so unreasonable.  Indeed, it is difficult to imagine how Lorentz symmetry 
can survive as a fundamental symmetry in a framework in which the space itself 
is viewed as an emergent property of the theory.  In string theory, quantum 
mechanics appears to be more fundamental than the symmetries of special of 
general relativity.  As a result, we adopt the perspective that Lorentz 
symmetry should appear as an emergent symmetry at long distances, but can be 
fundamentally absent at high energies.  

Despite being fundamentally nonrelativistic at short distances, our models 
of gravity with anisotropic scaling will describe propagating polarizations 
of the metric.  Restoring the explicit factors of the speed of light, the 
propagator for such gravitons will schematically take the form
\be
\label{prp}
\frac{1}{\omega^2-c^2\bk^2- G(\bk^2)^z},
\ee
where $G$ is a coupling constant.  (Generally, the denominator will also 
contain other powers of $\bk^2$ between 1 and $z$, which we omit here 
to keep this introductory discussion simple.)  

At high energies, the propagator is dominated by the anisotropic term 
$1/(\omega^2-G(\bk^2)^z)$.  The high-energy behavior of the theory is 
controlled by a free-field fixed point with anisotropic scaling.  For a 
suitably chosen $z$, this modification improves the short-distance behavior, 
shifting the critical dimension at which the theory is power-counting 
renormalizable.  The $c\bk^2$ term in (\ref{prp}) becomes important only at 
lower energies:  This term originates from a relevant deformation of the 
anisotropic UV fixed point, with $c$ a dimensionful coupling.  The 
propagator (\ref{prp}) is reproduced by the resummation of the high-energy 
propagator in the theory deformed by this relevant operator,
\be
\frac{1}{\omega^2-c^2\bk^2-G(\bk^2)^z}=
\frac{1}{\omega^2-G(\bk^2)^z}+\frac{1}{\omega^2-G(\bk^2)^z}\,c^2\bk^2\,
\frac{1}{\omega^2-G(\bk^2)^z}+\ldots.
\ee

At low energies, the theory naturally flows to $z=1$.  The relativistic 
scaling of space and time is ``accidentally restored,'' in the technical sense 
of renormalization theory.  In this low-energy regime, it is natural to adopt 
the perspective of a theory with relativistic scaling and absorb $c$ into the 
redefinition of the time coordinate, effectively setting $c=1$.   From the 
perspective of the $z=1$ IR fixed point, the higher-curvature terms which 
dominate the UV fixed point represent small corrections to the $z=1$ scaling, 
and the propagator (\ref{prp}) can be interpreted 
as 
\be
\frac{1}{\omega^2-\bk^2-G(\bk^2)^z}=
\frac{1}{\omega^2-\bk^2}+\frac{1}{\omega^2-\bk^2}\,G(\bk^2)^z\,
\frac{1}{\omega-\bk^2}+\ldots.
\ee
Unlike in relativistic higher-derivative theories 
mentioned above, higher time derivatives are not generated, and the usual 
problem of higher-derivative gravities with perturbative unitarity is 
eliminated.  

In this paper, we use these ideas to formulate a theory of gravity which would 
be power-counting renormalizable in $3+1$ dimensions.  Given the arguments 
above, this implies that $z=3$.  We develop the theory of gravity at such 
a ``$z$=3 Lifshitz point'' in Section~2.  Under the additional condition of 
``detailed balance,'' this theory turns out to be intimately related to 
topological gravity in three dimensions and the geometry of the Cotton 
tensor.  We discuss various properties of the $z=3$ UV fixed points, and study 
the relevant deformations which induce the flow at low energies to $z=1$, the 
value of the dynamical exponent in general relativity.  

In addition to $z=3$ gravity in $3+1$ dimensions and its infrared flow to 
$z=1$, in Section~\ref{secother} we briefly discuss the case of $z=4$ 
in $4+1$ and $3+1$ dimensions.  We also point out that another example of 
gravity with $z\neq 1$ has already appeared in the literature, under the name 
of ``ultralocal theory'' of gravity.  Section~\ref{secbb} contains a brief 
discussion of possible applications of anisotropic models of gravity in the 
context of AdS/CFT correspondence.

\section{Quantum Gravity in $3+1$ Dimensions at a $z=3$ Lifshitz Point}

Our aim is to construct a theory of gravity in $3+1$ dimensions with 
anisotropic scaling using the traditional framework of quantum field theory, 
\ie , path-intergral methods or canonical quantization.  Such an anisotropic 
theory of gravity, characterized by dynamical critical exponent 
$z=2$, was introduced in \cite{mqc}.  The main novelty of the present paper is 
that we are now interested in the case of $z=3$, which will lead to a 
power-counting renormalizable theory in $3+1$ dimensions.  Our construction 
parallels that of \cite{mqc}, which also contains additional details involving 
the general class of gravity models with anisotropic scaling.

\subsection{Fields, Scalings and Symmetries}

The quantum fields of our theory will include the spatial metric field 
$g_{ij}(\bx,t)$, which upon quantization describes propagating, interacting 
gravitons.   In this paper, we will define this theory on a fixed spacetime 
manifold $\CM$, and will not consider the possibility of summing over distinct 
topologies of spacetime.  On $\CM$, we will use coordinates 
\be
(t,\bx)\equiv (t,x^i),\qquad i=1,\ldots D,
\ee
with $D$ denoting the dimension of space.  For most of the paper, we will 
be interested in the case of $D=3$, but some of our arguments will be 
more instructive if we keep $D$ arbitrary.  Our notation throughout will be 
strictly nonrelativistic, unless stated otherwise.  For example, the covariant 
derivative $\nabla_i$ is defined with respect to the {\it spatial\/} metric 
$g_{ij}$, and we use 
$R^i{}_{jk\ell}$, $R_{ij}\equiv R^k{}_{ikj}$ and $R\equiv R^i_i$ to denote the 
Riemann tensor, the Ricci tensor and the Ricci scalar of the spatial 
metric $g_{ij}$ and its associated connection $\nabla_i$.    

\subsubsection{Anisotropic Scaling in Gravity}

The theory will be constructed so that it is compatible with anisotropic 
scaling with dynamical critical exponent $z$, 
\be
\label{scaling}
\bx\to b\bx,\qquad t\to b^zt.
\ee
In order for the theory to be power-counting renormalizable in $3+1$ spacetime 
dimensions, we will choose $z=3$, but for now we keep $z$ arbitrary.    

The scaling in (\ref{scaling}) is of course not diffeomorphism invariant (nor 
is it invariant under the gauge symmetries that we will impose on our system 
below), and should be interpreted in the following sense:  The theory will be 
designed such that it has a solution which describes an ultraviolet 
free-field fixed point with scaling properties given by (\ref{scaling}).  
At this fixed point, we will measure canonical dimensions of all objects in 
the units of spatial momenta.  In particular, the space-time anisotropy is 
reflected in the dimensions of time and space coordinates,
\be
[\bx]=-1,\qquad [t]=-z,
\ee
at this ultraviolet fixed point.

In addition to the spatial metric $g_{ij}$ (of signature $(+\ldots +)$), the 
field content of the theory will be given by a spatial vector $N_i$,  and a 
spatial scalar $N$.  The fields $N$ and $N_i$ are essentially the ``lapse'' 
and ``shift'' variables familiar from general relativity, where they appear in 
the process of the $3+1$ split of the four-dimensional spacetime metric.  
(The precise way in which these variables are related to the full spacetime 
metric can be found in \cite{mqc}.)  Using such ADM-like variables is 
particularly natural because of the fundamentally nonrelativistic nature of 
our system.  

In the case of general $z$, we postulate the classical scaling dimensions of 
the fields to be
\be
[g_{ij}]=0,\qquad [N_i]=z-1,\qquad [N]=0.
\ee
In the specific case $z=3$ of interest here, we have $[N_i]=2$, while $N$ and 
$g_{ij}$ are dimensionless.  

\subsubsection{Foliation-Preserving Diffeomorphisms}

In the anisotropic scaling (\ref{scaling}), the time dimension plays a 
privileged role.  We will encode this special role of time in the theory by 
assuming that in addition to being a differentiable manifold, our spacetime 
$\CM$ carries an extra structure -- that of a codimension-one foliation.%
\footnote{In differential geometry, a codimension-$q$ foliation $\CF$ on a 
$d$-dimensional manifold $\CM$ is defined as $\CM$ equipped with an atlas of 
coordinate systems $(y^a,x^i)$ $a=1,\ldots q$, $i=1,\ldots d-q$, such that the 
transition functions take the restricted form $(\tilde y^a,\tilde x^i)=
(\tilde y^a(y^b),\tilde x^i(y^b,x^j))$.  The general theory of foliations 
is reviewed \eg\ in \cite{lawson,godbillon,moerdijk}.}
This foliation structure $\CF$ is to be viewed as a part of the topological 
structure of $\CM$, before any notion of a Riemannian metric is introduced.  
The leaves of this foliation are the hypersurfaces of constant 
time.  Coordinate transformations adapted to the foliation are of the form
\be
\tilde x^i=\tilde x^i(x^j,t),\qquad \tilde t=\tilde t(t).
\ee
Thus, the transition functions are foliation-preserving diffeomorphisms.  We 
will denote the group of foliation-preserving diffeomorphisms of $\CM$ by 
$\dif(\CM)$.  In the local adapted coordinate system, the infinitesimal 
generators of $\dif(\CM)$ are given by
\be
\label{fdif}
\delta x^i=\zeta^i(t,\bx),\qquad \delta t=f(t).
\ee
We will simplify our presentation by further assuming that the spacetime 
foliation is topologically given by 
\be
\CM=\R\times\Sigma,
\ee
with all leaves of the foliation topologically equivalent to a fixed 
D-dimensional manifold $\Sigma$.  

Differential geometry of foliations is a well-developed branch of 
mathematics, and represents the proper mathematical setting for the class of 
gravity theories studied here.  We will not review the geometric theory of 
foliations in any detail here, instead referring the reader to 
\cite{lawson,godbillon,moerdijk}.  For example, there are two natural classes 
of functions that can be defined on a foliation:  In addition to functions 
that are allowed to depend on all coordinates, there is a special class of 
functions which take constant values on each leaf of the foliation.  We will 
call such functions ``projectable.''   

Foliations can be equipped with a Riemannian structure.  A Riemannian 
structure compatible with our codimension-one foliation of $\CM$ consists of 
three objects:  $g_{ij}$, $N_i$, and $N$, with $N$ a projectable function; 
both $N$ and $N_i$ transforms as vectors under the reparametrizations of 
time.  As pointed out above, these fields can be viewed as a decomposition of 
a Riemannian metric on $\CM$ into the metric $g_{ij}$ induced along the 
leaves, the shift variable $N_i$ and the lapse field $N$.  The generators 
of $\dif(\CM)$ act on the fields via
\bea
\label{foldif}
\delta g_{ij}&=&\p_i\zeta^kg_{jk}+\p_j\zeta^kg_{ik}+\zeta^k\p_kg_{ij}+f\,
\dot g_{ij},\nonumber\\
\delta N_i&=&\p_i\zeta^jN_j+\zeta^j\p_jN_i+\dot\zeta^jg_{ij}+\dot f\,N_i+f
\,\dot N_i,\\
\delta N&=&\zeta^j\p_jN+\dot f\,N+f\,\dot N.\nonumber
\eea
In \cite{mqc}, these transformation rules were derived by starting with the 
action of spacetime diffeomorphisms on the relativistic metric in the ADM 
decomposition, and taking the $c\to\infty$ limit.  We also saw in 
\cite{mqc}that $N_i$ and $N$ can be naturally interpreted as gauge 
fields associated with the time-dependent spatial diffeomophisms and the time 
reparametrizations, respectively.  In particular, since $N$ is the gauge 
field associated with the time reparametrization $f(t)$, it appears natural to 
restrict it to be a projectable function on the spacetime foliation $\CF$.  

If we wish instead to treat $N$ as an arbitrary function of spacetime, we have 
essentially two options.  First, we can allow an arbitrary spacetime-dependent 
$N$ as a background field, but integrate only over space-independent 
fluctuations of $N$ in the path integral.  As the second option, we will 
encounter situations in which $N$ {\it must\/} be allowed to be a general 
function of spacetime, because it participates in an additional gauge 
symmetry.  When that happens, we will integrate over the fluctuations of $N$ 
in the path integral.  An example of such an extra symmetry is the invariance 
under anisotropic Weyl transformations discussed in Section~\ref{secweyl} 
below, and in Section~5.2 of \cite{mqc}.

\subsection{Lagrangians}

We formally define our quantum field theory of gravity by a path 
integral,
\be
\int\CD g_{ij}\,\CD N_i\,\CD N\,\exp\{iS\}.  
\ee
Here $\CD g_{ij}\,\CD N_i\,\CD N$ denotes the path-integral measure whose 
proper treatment involves the Faddeev-Popov gauge fixing of the gauge symmetry 
$\dif(\CM)$, and $S$ is the most general action compatible with the 
requirements of gauge symmetry (and further restricted by unitarity).  As 
is often the case, this path integral is interpreted as the analytic 
continuation of the theory which has been Wick rotated to imaginary time 
$\tau=it$.  

Our next step is to construct the action $S$ compatible with our symmetry 
requirements.  
For simplicity, we will assume that all global topological effects can be 
ignored, freely dropping all total derivative terms and not discussing 
possible boundary terms in the action.  This is equivalent to assuming that 
our space $\Sigma$ is compact and its tangent bundle topologically trivial.  
The refinement of our construction which takes into account global topology 
and boundary terms is outside of the scope of the present work.  

\subsubsection{The Kinetic Term}

The kinetic term in the action will be given by the most general expression 
which is (i) quadratic in first time derivatives $\dot g_{ij}$ of the spatial 
metric, and (ii) invariant under the gauge symmetries of foliation-preserving 
diffeomorphisms $\dif(\CM)$.  The object that transforms covariantly under 
$\dif(\CM)$ is not $\dot g_{ij}$, but instead the second fundamental form 
\be
\label{kext}
K_{ij}=\frac{1}{2N}\left(\dot g_{ij}-\nabla_iN_j-\nabla_jN_i\right).     
\ee
This tensor measures the extrinsic curvature of the leaves of constant time in 
the spacetime foliation $\CF$.   In terms of $K_{ij}$ and its trace 
$K\equiv g^{ij}K_{ij}$, the kinetic term is given by
\be
\label{kin}
S_K=\frac{2}{\kappa^2}\int dt\,d^D\bx\,\sqrt{g}N\left(K_{ij}K^{ij}
-\lambda K^2\right).
\ee
This kinetic term contains two coupling constants: $\kappa$ and $\lambda$.  
The dimension of $\kappa$ depends on the spatial dimension $D$:  Since the 
dimension of the volume element is 
\be
\label{dimint}
[dt\,d^D\bx]=-D-z,
\ee
and each time derivative contributes $[\p_t]=z$, the scaling dimension of 
$\kappa$ is
\be
\label{dimk}
[\kappa]=\frac{z-D}{2}.
\ee
As intended, this coupling will be dimensionless in $3+1$ spacetime dimensions
if $z=3$.  

The presence of an additional, dimensionless coupling $\lambda$ reflects the 
fact that each of the two terms in (\ref{kin}) is separately invariant under 
$\dif(\CM)$.  In other words, the requirement of $\dif(\CM)$ symmetry allows 
the generalized De~Witt ``metric on the space of metrics'' 
\be
\label{dwm}
G^{ijk\ell}=\frac{1}{2}\left(g^{ik}g^{j\ell}+g^{i\ell}g^{jk}\right)-\lambda 
g^{ij}g^{k\ell}
\ee
to contain a free parameter $\lambda$.  It is this generalized De~Witt metric 
that defines the form quadratic in $K_{ij}$ which appears in the kinetic term 
(see \cite{mqc}).  

In general relativity, the requirement of invariance under 
all spacetime diffeomophisms forces $\lambda=1$.  In our theory with 
$\dif(\CM)$ gauge invariance, $\lambda$ represents a dynamical coupling 
constant, susceptible to quantum corrections.  

It is interesting to note that the kinetic term $S_K$ is universal, 
and independent of both the desired value of $z$ and the dimension of 
spacetime.  The only place where the value of $z$ shows up in $S_K$ is in the 
scaling dimension of the integration measure (\ref{dimint}), which in turn 
determines the dimension (\ref{dimk}) of $\kappa$.  The main difference 
between theories with different $z$ will be in the pieces of the action which 
are independent of time derivatives.  
  
\subsubsection{The Potential}

The logic of effective field theory suggests that the complete action should 
contain all terms compatible with the imposed symmetries, which are of 
dimension equal to or less than the dimension of the kinetic term, 
$[K_{ij}K^{ij}]=2z$.  
In addition to $S_K$, which contains the two independent terms of second order 
in the time derivatives of the metric, the general action will also contain 
terms that are independent of time derivatives.  Since our framework is 
fundamentally non-relativistic, we will refer to all terms in the action 
which are independent of the time derivatives (but do depend on spatial 
derivatives) simply as the ``potential.''

There is a simple way how to construct potential terms invariant under 
our gauge symmetry $\dif(\CM)$:  Starting with any scalar function 
$V[g_{ij}]$ which depends only on the metric and its spatial derivatives, 
the following potential term
\be
S_V=\int dt\,d^D\bx\,\sqrt{g}N\,V[g_{ij}]
\ee
will be invariant under $\dif(\CM)$.  

Throughout this paper, our strategy is to focus first on the potential terms 
of the same dimension as $[K_{ij}K^{ij}]$, at first ignoring all possible 
relevant terms of lower dimensions in $V$.  This is equivalent to focusing 
first on the high-energy limit, where such highest-dimension terms dominate.  
Once the high-energy behavior of the theory is understood, one can restore the 
relevant terms, and study the flows of the theory away from the UV fixed point 
that such relevant operators induce in the infrared. 

With our choice of $D=3$ and $z=3$, there are many examples of terms in $V$ of 
the same dimension as the kinetic term in (\ref{kin}).  Some such terms are 
quadratic in curvature, 
\be
\nabla_kR_{ij}\nabla^kR^{ij},\qquad \nabla_kR_{ij}\nabla^iR^{jk},\qquad 
R\Delta R,\qquad R^{ij}\Delta R_{ij};
\ee
they will not only add interactions but also modify the propagator.  Other 
terms, such as 
\be
R^3,\qquad R^i_jR^j_kR^k_i,\qquad RR_{ij}R^{ij},
\ee
are cubic in curvature, and therefore represent pure interacting terms.  
Some of the terms of the correct dimension are related by the Bianchi 
identity and other symmetries of the Riemann tensor, or differ only up to a 
total derivative.   Additional constraints on the possible values of the 
couplings will likely follow from the requirements of stability and unitarity 
of the quantum theory.  However, the list of independent operators appears to 
be prohibitively large, implying a proliferation of couplings which makes 
explicit calculations rather impractical.  

\subsection{UV Theory with Detailed Balance}

In order to reduce the number of independent coupling constants, we will 
impose an additional symmetry on the theory.  The reason for this restriction 
is purely pragmatic, to limit the proliferation of independent couplings 
mentioned in the previous paragraph.  The way in which this restriction will 
be implemented, however, is very reminiscent of methods used in nonequilibrium 
critical phenomena and quantum critical systems.  As a result, it is natural 
to suspect that there might also be conceptual reasons behind restricting the 
general class of classical theories to conform to this framework in systems 
with gravity as well.

We will require the potential term to be of a special form,
\be
\label{dbc}
S_V=\frac{\kappa^2}{8}\int dt\,d^D\bx\,\sqrt{g}N\,E^{ij}\CG_{ijk\ell} 
E^{k\ell},
\ee
and will further demand that $E^{ij}$ itself follow from a variational 
principle,
\be
\label{eom}
\sqrt{g}E^{ij}=\frac{\delta W[g_{k\ell}]}{\delta g_{ij}}
\ee
for some action $W$.  The two copies of $E^{ij}$ in (\ref{dbc}) are contracted 
by $\CG_{ijk\ell}$, the inverse of the De~Witt metric (\ref{dwm}).  
Loosely borrowing terminology from nonequilibrium dynamics, we will say that 
theories whose potential is of the form (\ref{dbc}) with (\ref{eom}) for some 
$W$ satisfy the ``detailed balance condition.''  

In the context of condensed matter, the virtue of the detailed balance 
condition is in the simplification of the renormalization properties.  
Systems which satisfy the detailed balance condition with some $D$-dimensional 
action $W$ typically exhibit a simpler quantum behavior than a generic 
theory in $D+1$ dimensions.  Their renormalization can be reduced to the 
simpler renormalization of the associated theory described by $W$, followed 
by one additional step -- the renormalization of the relative couplings 
between the kinetic and potential terms in $S$.  Examples of this phenomenon 
include scalar fields \cite{zinn} or Yang-Mills gauge theories 
\cite{zinnzwan,cym}.  Investigating the precise circumstances under which 
this ``quantum inheritance principle'' holds for gravity systems will be 
important for understanding the quantum properties of gravity models with 
nonrelativistic values of $z$.   

Since we are primarily interested in theories which are 
spatially isotropic, $W$ must be the action of a relativistic theory in 
Euclidean signature.  (Obvious generalizations to theories with additional 
spatial anisotropies are clearly possible, but will not be pursued in this 
paper.)  In \cite{mqc}, a theory of gravity in $D+1$ dimensions satisfying the 
detailed balance condition was constructed, with $W$ the Einstein-Hilbert 
action
\be
\label{ehact}
W=\frac{1}{\kappa_W^2}\int d^D\bx\,\sqrt{g}(R-2\lambdaw).
\ee
The potential $S_V$ of this theory takes the form 
\be
\label{pottwo}
S_V=\frac{\kappa^2}{8\kappa_W^4}\int dt\,d^D\bx\,\sqrt{g}N
\left(R^{ij}-\frac{1}{2}Rg^{ij}+\lambdaw g^{ij}\right)
\CG_{ijk\ell} 
\left(R^{k\ell}-\frac{1}{2}Rg^{k\ell}+\lambdaw g^{k\ell}\right).
\ee
At short distances, the curvature term in $W$ dominates over $\lambdaw$, 
and the resulting potential $S_V$ is quadratic in the curvature tensor: The 
theory exhibits anisotropic scaling with $z=2$ in the UV\@.  Turning on 
$\lambdaw$ in $W$ leads to lower-dimension terms in $S_V$ which dominate at 
long distances, and the theory undergoes a classical flow to $z=1$ in the 
IR\@.  The anisotropic scaling in the UV shifts the critical dimension of this 
theory, which is now renormalizable by power counting in $2+1$ dimensions.  
In dimensions higher than $2+1$, the theory with potential (\ref{pottwo}) 
is merely a low-energy effective field theory, and can be expected to break 
down at the scale set by the dimensionful coupling $\kappa_W^{}$.  

Here we are interested in constructing a theory which satisfies detailed 
balance, and exhibits the short-distance scaling with $z=3$ leading to 
power-counting renormalizability in $3+1$ dimensions.  Therefore, $E^{ij}$ 
must be of third order in spatial derivatives.  As it turns out, there is a 
unique candidate for such an object: the Cotton tensor
\be
\label{cotton}
C^{ij}=\varepsilon^{ik\ell}\nabla_k\left(R^j_\ell-\frac{1}{4}R\delta^j_\ell
\right).
\ee
This tensor not only exhibits all the required symmetries, it also follows 
from a variational principle. 

\subsubsection{Properties of the Cotton Tensor}

The Cotton tensor enjoys several symmetry properties which may not be 
immediately obvious from its definition in (\ref{cotton}):

(i) It is symmetric and traceless,
\be
C^{ij}=C^{ji},\qquad g_{ij}C^{ij}=0.
\ee 

(ii) It is transverse (or covariantly conserved),
\be
\label{transv}
\nabla_iC^{ij}=0.
\ee

(iii) It is conformal, with conformal weight $-5/2$.  More precisely, 
under local spatial Weyl transformations 
\be
g_{ij}\to\exp\left\{2\Omega(\bx)\right\}g_{ij},
\ee
it transforms as 
\be
\label{cotconf}
C^{ij}\to\exp\left\{-5\Omega(\bx)\right\}C^{ij},
\ee
with no terms containing derivatives of $\Omega(\bx)$.  

The Cotton tensor plays an important role in geometry.  Recall that in 
dimensions $D>3$, the property of conformal flatness of a Riemannian metric 
is equivalent to the vanishing of the Weyl tensor $C_{ijk\ell}$, defined 
as the completely traceless part of the Riemann tensor:
\bea
C_{ijk\ell}=R_{ijk\ell}&-&\frac{1}{D-2}\left(g_{ik}R_{j\ell}-g_{i\ell}R_{jk}-
g_{jk}R_{i\ell}+g_{j\ell}R_{ik}
\right)\cr
&&\qquad{}+\frac{1}{(D-1)(D-2)}\left(g_{ik}g_{j\ell}
-g_{i\ell}g_{jk}\right)R.
\eea
In $D=3$, however, the Weyl tensor vanishes identically, and another object 
has to take over the role in the criterion of conformal flatness of 
3-manifolds.  This object is the Cotton tensor, of third order in spatial 
derivatives.    

The Cotton tensor also plays an important role in physics.  In the initial 
value problem of the Hamiltonian formulation of general relativity, it is 
natural to ask what set of initial conditions can be freely specified for 
the metric and its canonical momenta, without violating the constraint part 
of Einstein's equations.  It was shown by York \cite{york1,york2,york3} that 
the Cotton tensor plays a central role in answering this question.  The 
correct initial conditions are set by specifying the values of two 
tensors with the symmetries of the Cotton tensor:  One related to the 
initial value for the conformal structure of the spatial metric, and the other 
specifying the initial value of the conjugate momenta.  For this 
reason, $C^{ij}$ is often referred to as the ``Cotton-York tensor'' in the 
physics literature.  

Lastly, the Cotton tensor follows from a variational principle, with action
\be
\label{gcs}
W=\frac{1}{w^2}\int_\Sigma\omega_3(\Gamma).  
\ee
Here $w^2$ is a dimensionless coupling, and 
\be
\omega_3(\Gamma)=\trace\left(\Gamma\wedge d\Gamma+\frac{2}{3}\Gamma\wedge\Gamma
\wedge\Gamma\right)\equiv\varepsilon^{ijk}\left(\Gamma^{m}_{i\ell}\p_j
\Gamma^{\ell}_{km}+\frac{2}{3}\Gamma^{n}_{i\ell}\Gamma^{\ell}_{jm}
\Gamma^{m}_{kn}\right)d^3\bx
\ee
is the gravitational Chern-Simons term, with the Christoffel symbols 
$\Gamma^{i}_{jk}$ treated as known functionals of the metric $g_{ij}$, and 
not as independent variables.  The variation of (\ref{gcs}) with respect to 
$g_{ij}$ yields the vanishing of the Cotton tensor as the equations of 
motion.  

Without any loss of generality, we will assume that the coupling $w^2$ is 
positive; its sign can be changed by flipping the orientation of the 
3-manifold $\Sigma$.  Unlike in Chern-Simons gauge theories with a compact 
gauge group, the coupling constant of Chern-Simons gravity in $2+1$ dimensions 
is not quantized, as a result of the absence of large gauge tranformations.  
In our framework, however, we are only interested in the action of a theory in 
three dimensions in ``imaginary time,''  and require that this Euclidean 
action be real.  This is to be contrasted with the conventional interpretation 
of the three-dimensional theory, which involves analytic continuation to real 
time in $2+1$ dimensions, and imposes a slightly different reality conditions 
on the action.

\subsubsection{$z=3$ Gravity with Detailed Balance}

Having reviewed some of the properties of the Cotton tensor, we can now write 
down the full action of our $z=3$ gravity theory in $3+1$ dimensions: 
\bea
\label{fuact}
S&=&\int dt\,d^3\bx\,\sqrt{g}\,N\left\{\frac{2}{\kappa^2}\left(
K_{ij}K^{ij}-\lambda K^2\right)-\frac{\kappa^2}{2w^4}C_{ij}C^{ij}
\right\}\cr
&&{}=\int dt\,d^3\bx\,\sqrt{g}\,N\left\{\frac{2}{\kappa^2}\left(
K_{ij}K^{ij}-\lambda K^2\right)\right.\cr
&&\qquad\qquad\qquad\left.{}-\frac{\kappa^2}{2w^4}\left(\nabla_iR_{jk}
\nabla^iR^{jk}-\nabla_iR_{jk}\nabla^jR^{ik}-\frac{1}{8}\nabla_iR\nabla^iR
\right)\right\}.
\eea
As a result of the uniqueness of the Cotton tensor, the action given in 
(\ref{fuact}) describes the most general $z=3$ gravity satisfying the detailed 
balance condition, modulo the possible addition of relevant terms, which will 
be discussed in Section~\ref{secrede}.  

We can demonstrate that after the Wick rotation to imaginary time, this  
action can be written -- up to a total derivative -- as a sum of squares, 
\bea
\label{imact}
S&=&i\int d\tau\,d^3\bx\,\sqrt{g}\,N\left\{\frac{2}{\kappa^2}\left(
K_{ij}K^{ij}+\lambda K^2\right)+\frac{\kappa^2}{2w^4}C_{ij}C^{ij}
\right\}\cr
&&{}=2i\int d\tau\,d^3\bx\,\sqrt{g}\,N\left(\frac{1}{\kappa}K_{ij}
-\frac{\kappa}{2w^2}C_{ij}\right)G^{ijk\ell}\left(\frac{1}{\kappa}K_{k\ell}
-\frac{\kappa}{2w^2}C_{k\ell}\right),
\eea
First, $C_{ij}G^{ijk\ell}C_{k\ell}=C_{ij}C^{ij}$ because $C^{ij}$ is 
traceless.  As to the cross-terms $K_{ij}G^{ijk\ell}C_{k\ell}$, they can be 
written as a total derivative,
\bea
&&\qquad\frac{1}{w^2}\int d\tau\,d^3\bx\,\sqrt{g}\,N\,K_{ij}G^{ijk\ell}
C_{k\ell}=\frac{1}{2w^2}\int d\tau\,d^3\bx\,\sqrt{g}\,\left(\dot g_{ij}
-\nabla_iN_j-\nabla_jN_i\right)C_{ij}\cr
&&=\int d\tau\,d^3\bx\,\sqrt{g}\left(\dot g_{ij}
\frac{\delta W}{\delta g_{ij}}+\frac{1}{w^2}\nabla_i\left(N_jC^{ij}\right)
\right)=\int d\tau\,d^3\bx\left(\dot \CL+\frac{1}{w^2}\p_i\left(
\sqrt{g}N_jC^{ij}\right)\right),\nonumber
\eea
where we used the transverse property (\ref{transv}) of $C_{ij}$, and 
$\CL$ is the Lagrangian density of the action $W$ in (\ref{gcs}).    

Introducing an auxiliary field $B^{ij}$, it is conventient to rewrite the 
imaginary-time action as
\be
\label{bact}
S=2i\int d\tau\,d^3\bx\,\sqrt{g}\,N\left\{B^{ij}\left(\frac{1}{\kappa}K_{ij}
-\frac{\kappa}{2w^2}C_{ij}\right)
-B^{ij}\CG_{ijk\ell}B^{k\ell}\right\}.
\ee
This form of the action, with all terms at least linear in the auxiliary 
field $B^{ij}$ and with the linear term proportional to a gradient flow 
equation, is symptomatic of theories satisfying the detailed balance condition 
in the context of condensed matter systems, in particular in the theory of 
quantum and dynamical critical phenomena \cite{halperin,ma}, stochastic 
quantization \cite{parisi,namiki}, and nonequilibrium statitistical 
mechanics \cite{lebellac}.  

In that condensed-matter context, the property of detailed balance often has 
one interesting implication.  If a quantum critical system in $D+1$ dimensions 
satisfies detailed balance with some $W$ in $D$ dimensions, the partition 
function of the theory described by $W$ yields a natural solution of the 
Schr\"odinger equation of the theory in $D+1$ dimensions, which plays the role 
of a candidate ground-state wavefunction.  Similarly, in nonequilibrium 
statistical mechanics and dynamical critical phenomena, the corresponding 
statement is essentially the Wick rotation of this correspondence to imaginary 
time:  The partition function of the $D$ dimensional theory defined by $W$ 
represents an equilibrium state solution of the dynamical theory with 
detailed balance in $D+1$ dimensions.  

In our case, this correspondence formally suggests that 
\be
\label{nowave}
\Psi_0[g_{ij}(\bx)]=\exp\left\{-\frac{1}{2w^2}\int\trace\left(\Gamma\wedge 
d\Gamma+\frac{2}{3}\Gamma\wedge\Gamma\wedge\Gamma\right)\right\},
\ee
is a solution of the Schr\"odinger equation of the theory in canonical 
quantization.  One might be tempted to consider (\ref{nowave}) a candidate for 
the ground state wavefunction of quantum gravity with $z=3$.  However, it 
becomes quickly obvious that (\ref{nowave}) is an unphysical solution:  $W$ 
is not bounded from below, $\Psi_0$ is non-normalizable, and any attempts to 
build a spectrum of excited states above this hypothetical ground state 
lead inevitably to pathologies.  

This is to be compared to relativistic Yang-Mills gauge theory in $3+1$ 
dimensions, which is in fact surprisingly similar to our $z=3$ theory of 
gravity, in at least two respects:  

(i) it also satisfies detailed balance, 

(ii) the corresponding action $W$ in three dimensions is also given by the 
Chern-Simons action, $\int\omega_3(A)$, with $A$ the Yang-Mills one-form gauge 
field.  

Similarly to (\ref{nowave}), the candidate ground-state wavefunction 
\be
\label{nowaveym}
\Psi_0[A]\sim\exp\left\{-\int\omega_3(A)\right\}
\ee
is formally a solution of the Schr\"odinger equation for Yang-Mills theory in 
$3+1$ dimensions, but an equally unphysical one.  A very clear and conclusive 
analysis showing why (\ref{nowaveym}) is unphysical can be found in 
\cite{witten}.%
\footnote{In contrast, for some theories with detailed balance, 
$\Psi_0\sim\exp\{-W/2\}$ does represent a physical normalizable ground-state 
wavefunction.  Examples include the Lifshitz scalar theory (as discussed for 
example in \cite{ardonne}), and the quantum critical Yang-Mills with $z=2$ in 
$4+1$ dimensions \cite{cym}.}
The fate of the formal solution (\ref{nowave}) of the gravity theory is the 
same as the fate of (\ref{nowaveym}) in Yang-Mills:  The failure for 
(\ref{nowave}) to be the true ground-state wavefuction is not a flaw of the 
theory, it just means that -- just as in $3+1$ dimensional Yang-Mills theory 
-- the true ground-state wavefunction is much harder to find.  

In passing, it is amusing to note that essentially the same expression 
$\Psi_0$ given in (\ref{nowave}) was proposed some time ago as a candidate 
ground-state wavefunction of loop quantum gravity, where it is known as 
the ``Kodama wavefunction.''  Again, a long list of conclusive reasons why 
this cannot possibly be the physical wavefunction of quantum gravity can be 
found in \cite{witten}. 

\subsubsection{Anisotropic Weyl Invariance at $\lambda=1/3$}
\label{secweyl}

The fact that the Cotton tensor is conformal suggests that, under special 
circumstances, the classical action of $z=3$ gravity in $3+1$ may be invariant 
under suitably defined local scale transformations.  As we now show, this is 
indeed the case:  With $\lambda=1/3$, our $z=3$ theory develops a classical 
anisotropic Weyl invariance, similar to that observed in \cite{mqc} in the 
case of the $z=2$ theory in $2+1$ dimensions with $\lambda=1/2$.  

To see that, we decompose the metric by pulling out the overall scale factor,  
\be 
g_{ij}=g^{1/3}\tilde g_{ij}=e^\phi\tilde g_{ij},\qquad N_i=e^\phi\tilde N_i,
\ee
where $\det\tilde g_{ij}=1$. With this decomposition, the $z=3$ action 
(\ref{fuact}) becomes
\bea
\label{actionw}
S&=&\frac{1}{2}\int dt\,d^3\bx\left\{\frac{e^{3\phi/2}}{\kappa^2N}\left[
\left(\dot{\tilde g}_{ij}-\tilde\nabla_i\tilde N_j-\tilde\nabla_j\tilde N_i
\right)
(\tilde g^{ik}\tilde g^{j\ell}-\lambda\tilde g^{ij}\tilde g^{k\ell})
\left(\dot{\tilde g}_{k\ell}-\tilde\nabla_k\tilde N_\ell-\tilde\nabla_\ell
\tilde N_k\right)\right.\right.\cr
&&\left.\qquad{}+3(1-3\lambda)\left(\dot\phi-\tilde g^{ij}\tilde N_i\p_j\phi
\right)^2-4(1-3\lambda)\tilde\nabla_i\tilde N_j\,\tilde g^{ij}
\left(\dot\phi-\tilde g^{ij}\tilde N_i\p_j\phi\right)\right]\cr
&&\left.\qquad\qquad\qquad\qquad\qquad{}-\frac{\kappa^2}{w^4}
\frac{N}{e^{3\phi/2}}\tilde C^{ij}\tilde g_{ik}\tilde g_{j\ell}\tilde 
C^{k\ell}\right\},
\eea
where $\tilde\nabla_i$ is the covariant derivative associated with 
$\tilde g_{ij}$, $\tilde C^{ij}$ is the Cotton tensor of $\tilde g_{ij}$, 
and we used the conformal property $\tilde C^{ij}=e^{5\phi/2}C^{ij}$ of the 
Cotton tensor which follows from (\ref{cotconf}).  

By inspection, the action (\ref{actionw}) will be invariant under local Weyl 
transformations of the metric 
\be
\label{aniswi}
g_{ij}\to\exp\left\{2\Omega(t,\bx)\right\}g_{ij},
\ee
if we allow the Weyl rescalings to act on $N$ and $N_i$ by
\be
\label{aniswii}
N\to\exp\left\{3\Omega(t,\bx)\right\}N,\qquad 
N_i\to\exp\left\{2\Omega(t,\bx)\right\}N_i,
\ee
and if we also set $\lambda=1/3$.  This conformal choice of $\lambda$ 
eliminates all terms with derivatives of $\phi$ in (\ref{actionw}).  
Note that with the addition of the new gauge symmetry 
(\ref{aniswi}-\ref{aniswii}) to $\dif(\CM)$, the lapse field $N$ can no longer 
be a projectable function on the foliation, and must be allowed to depend on 
$x^i$ as well.  

It is reassuring to find that the spacetime-dependent anisotropic Weyl 
transformations (\ref{aniswi}) and (\ref{aniswii}) in fact represent the local 
version of the rigid anisotropic scaling (\ref{scaling}) with dynamical 
exponent $z=3$.  To see that, recall that $N$, $N_i$ and $g_{ij}$ can be 
reassembled into a spacetime metric in $3+1$ dimensions, with $g_{00}\sim 
-N^2$.  The scaling rules (\ref{aniswii}) that we found by requiring the Weyl 
invariance of the kinetic term then imply 
$g_{00}\to\exp\left\{6\Omega(t,\bx)\right\}g_{00}$.  In the flat background 
given by $N=1$, $N_i=0$ and $g_{ij}$ the flat Euclidean metric, the Weyl 
transformations with constant $\Omega$ reduce precisely to the anisotropic 
scaling (\ref{scaling}) with the value of the dynamical critical exponent 
$z=3$, which was the starting point of our construction of gravity at a $z=3$ 
Lifshitz point. 

Quantum corrections can be expected to generate violations of local 
anisotropic Weyl invariance.  Lessons from relativistic models suggest that 
such conformal anomalies vanish in theories with a sufficient degree of 
supersymmetry, and it should be interesting to investigate the conditions 
which lead to similar cancellations of conformal anomalies in the 
nonrelativistic models.  

\subsection{At the Free-Field Fixed Point}

The action (\ref{fuact}) of the $z=3$ theory with detailed balance contains 
three dimensionless coupling constants: $\kappa$, $\lambda$ and $w$.  
However, only one of them, $w$, controls the strength of interactions.  
The noninteracting limit corresponds to sending $w\to 0$, while keeping 
$\lambda$ and the ratio
\be
\gamma=\frac{\kappa}{w}
\ee
fixed.  This limit yields a two-parameter family of free-field fixed 
points, parametrized by $\lambda$ and $\gamma$.  

In preparation for the study of the full interacting theory, it is useful to 
first investigate the properties of this family of free-field fixed points.   
The linearization of the $z=3$ theory is performed in exactly the same way as 
in \cite{mqc} for the analogous case of the $z=2$ gravity and we will 
therefore be relatively brief, referring the reader to \cite{mqc} for further 
details.  

We expand the theory in small fluctuations $h_{ij}$, $n$ and $n_i$ around the 
flat background,
\be
g_{ij}\approx\delta_{ij}+wh_{ij},\qquad N\approx1+wn,\qquad N_i\approx wn_i.
\ee
The reference background is a solution of the equations of motion of the 
$z=3$ theory (\ref{fuact}).  
Keeping only quadratic terms in the action, $n$ drops out from the theory.  
A natural gauge choice is
\be
\label{fgf}
n_i=0.
\ee 
This fixes most of the $\dif(\CM)$ gauge symmetry, leaving time-independent 
spatial diffeomorphisms ${\rm Diff}(\Sigma)$ unfixed.  The residual 
${\rm Diff}(\Sigma)$ gauge symmetry can be conventiently fixed by setting 
\be
\label{resgf}
\p_ih_{ij}-\lambda\p_j h=0,
\ee
where $h\equiv h_{ii}$.  Imposing this condition at some fixed time slice 
$t=t_0$ effectively fixes the residual ${\rm Diff}(\Sigma)$ invariance.  The 
Gauss constraint
\be
\p_i\dot h_{ij}-\lambda\p_j \dot h=0
\ee
(which follows from varying the original action with respect to $n_i$) then 
ensures that (\ref{resgf}) stays valid at all times.  

In order to diagonalize the linearized equations of motion and read off the 
dispersion relation of the propagating modes implied by our gauge choice 
(\ref{fgf}) and (\ref{resgf}), it is convenient to first redefine the 
variables by introducing 
\be
H_{ij}\equiv h_{ij}-\lambda\delta_{ij}h; 
\ee
the gauge condition (\ref{resgf}) implies that $H_{ij}$ is transverse.  
We then decompose the transverse tensor $H_{ij}$ into its transverse 
traceless part $\tilde H_{ij}$ and its trace $H$, 
\be
H_{ij}=\tilde H_{ij}+\frac{1}{2}\left(\delta_{ij}-
\frac{\p_i\p_j}{\p^2}\right)H.
\ee
This choice of variables diagonalizes the equations of motion in our 
gauge.  Since the kinetic term is universal, its analysis in the $z=3$ theory 
is identical to that presented for $z=2$ in Section~4.5 of\cite{mqc}.  In   
our gauge and in terms of the new variables, the kinetic term takes the form
\be
\label{linkin}
S_K\approx\frac{1}{2\gamma^2}\int dt\,d^3\bx\left\{\dot{\tilde H}_{ij}
\dot{\tilde H}_{ij}+\frac{1-\lambda}{2(1-3\lambda)}\dot H^2\right\}.
\ee  
It would appear that the dependence of the kinetic term of $H$ on $\lambda$ 
can be absorbed into a rescaling of $H$, but we choose not to do so, because 
it would obscure the geometric origin of $H$ in the full nonlinear theory.  

On the other hand, the potential term of the $z=3$ theory reduces to
\be
\label{linpot}
S_V\approx-\frac{\gamma^2}{8}\int dt\,d^3\bx\,\tilde H_{ij}(\p^2)^3\tilde 
H_{ij}.
\ee
Because of the conformal properties of the Cotton tensor, the potential 
term in the Gaussian approximation depends only on $\tilde H_{ij}$ and not 
on $H$.  

As pointed out in \cite{mqc}, the kinetic term (\ref{linkin}) indicates that 
two values of $\lambda$ play a special role.  At $\lambda=1/3$, the theory 
becomes compatible with the local anisotropic Weyl invariance discussed 
in Section~\ref{secweyl} above.  At that value of $\lambda$, the scalar mode 
$H$ is a gauge artifact.  The kinetic term for $H$ also appears singular at 
$\lambda=1$.  As explained in \cite{mqc}, this happens because at this special 
value of $\lambda$, the linearized theory exhibits an extra gauge invariance, 
which can be used to eliminate physical excitations of $H$ as well.  

The transverse traceless tensor $\tilde H_{ij}$ contains two propagating 
physical polarizations.  These gravitons satisfy a nonrelativistic gapless 
dispersion relation,
\be
\omega^2=\frac{\gamma^4}{4}(\bk^2)^3.  
\ee

For values of $\lambda$ outside of the two special values 1 and $1/3$, the 
scalar mode $H$ will represent a physical degree of freedom, with its 
linearized equation of motion given simply by $\ddot H=0$.  When the theory is 
deformed by relevant operators, the equation of motion for $H$ will contain 
terms with spatial derivatives up to fourth order, which is not enough to 
yield a propagator with good ultraviolet properties.  It appears that in order 
to make the theory power-counting renormalizable at generic values of 
$\lambda$ not equal to 1 or $1/3$, either the scalar mode would have to be 
eliminated by an extra gauge symmetry, or superrenormalizable terms which give 
short-distance spatial dynamics to the scalar mode need to be added to the 
potential.  We will briefly return to this point in Section~\ref{secsuper}.

\subsection{Relevant Deformations and the Infrared Flow to $z=1$}
\label{secrede}

So far we have concentrated on terms of the highest dimension terms in $S$.  
These terms will dominate the short-distance dynamics.  At long distances, 
relevant deformations by operators of lower dimensions will become important, 
in addition to the RG flows of the dimensionless couplings.  

One could relax the condition of detailed balance, and simply ask that 
the action $S$ in $3+1$ dimensions be a general combination of all marginal 
and relevant terms.  The action of the theory would then take the form
\be
S=\int dt\,d^3\bx\,\sqrt{g}\sum_{[\CO_J]=6}\lambda_J\CO_J+\int dt\,d^3\bx\,
\sqrt{g}\sum_{[\CO_A]<6}\lambda_A\CO_A,
\ee
where the index $J$ goes over all independent marginal terms compatible with 
$\dif(\CM)$, while $A$ parametrizes all independent relevant operators 
compatible with this symmetry.  $\lambda_J$ and $\lambda_A$ are the 
corresponding coupling constants.    

It would be desirable to analyze the quantum properties, in particular 
the RG flow patterns, of this general family of models.  However, the 
proliferation of operators with dimensions ${}\leq 6$ makes this analysis 
difficult, and we will again resort to theories which satisfy the additional 
property of detailed balance.    

\subsubsection{Relevant Deformations with Detailed Balance}

In order for the deformed theory to satisfy the detailed balance condition, 
the relevant deformations themselves must originate from an action principle 
in $D$ dimensions, subjected to the requirement of diffeomorphism invariance.  
Adding all possible relevant terms to the Chern-Simons action (\ref{gcs}), we 
get
\be
\label{cseh}
W=\frac{1}{w^2}\int\omega_3(\Gamma)+\mu\int d^3\bx\,
\sqrt{g}(R-2\lambdaw).
\ee
This is essentially the action of topologically massive gravity 
\cite{djt1,djt2}, a theory which has been argued to be renormalizable 
\cite{deser,keszthelyi} and possibly finite.%
\footnote{The main difference between (\ref{cseh}) and topologically 
massive gravity stems from the fact that here we are only interested in the 
Euclidean-signature version of (\ref{cseh}), with the real action $W$.  
In topologically massive gravity, the Euclidean action $W$ is interpreted as 
the Wick rotation of the real action from the physical signature $2+1$, 
leading to a slightly different reality condition on $W$, with $w^2$ purely 
imaginary.  There has been a recent resurgence of interest in topological 
massive gravity, initiated by \cite{lss}; see also \cite{witten3}.}
The coupling constants $\mu$ and $\lambdaw$ are of dimension 
$[\mu]=1$ and $[\lambdaw]=2$.  

The relevant operators in the action $W$ of (\ref{cseh}) induce relevant terms 
in the potential term $S_V$ of our $z=3$ theory.  The full action in $3+1$ 
dimensions which satisfies detailed balance with respect to (\ref{cseh}) is 
given by
\bea
\label{defpt}
&&S=\int dt\,d^3\bx\,\sqrt{g}\,N\left\{\frac{2}{\kappa^2}
K_{ij}G^{ijk\ell}K_{k\ell}-\frac{\kappa^2}{2}\left[\frac{1}{w^2}C^{ij}
-\frac{\mu}{2}\left(R^{ij}-\frac{1}{2}Rg^{ij}+\lambdaw g^{ij}\right)\right]
\right.\cr
&&\qquad\qquad\qquad\qquad\left.{}\times\CG_{ijk\ell}
\left[\frac{1}{w^2}C^{k\ell}-\frac{\mu}{2}\left(R^{k\ell}-\frac{1}{2}Rg^{k\ell}
+\lambdaw g^{k\ell}\right)\right]\right\}.
\eea
It is useful to organize the terms in (\ref{defpt}) in the order of their 
descending dimensions, 
\bea
\label{dfpt}
&&S=\int dt\,d^3\bx\,\sqrt{g}\,N\left\{\frac{2}{\kappa^2}
\left(K_{ij}K^{ij}-\lambda K^2\right)-\frac{\kappa^2}{2w^4}C_{ij}C^{ij}
+\frac{\kappa^2\mu}{2w^2}\varepsilon^{ijk}R_{i\ell}\nabla_jR_k^{\ell}\right.\cr
&&\qquad\qquad\qquad\left.{}-\frac{\kappa^2\mu^2}{8}R_{ij}R^{ij}
+\frac{\kappa^2\mu^2}{8(1-3\lambda)}\left(
\frac{1-4\lambda}{4}R^2+\lambdaw R-3\Lambda_W^2\right)\right\}.
\eea
At long distances, the potential is dominated by the last two terms in 
(\ref{dfpt}): the spatial curvature scalar and the constant term.  These 
leading terms in the potential combine with the kinetic term, and as a result, 
the theory flows in the infrared to $z=1$.  

This infrared limit of the deformed theory should be compared to general 
relativity.  As is well-known, the Einstein-Hilbert action in $3+1$ dimensions 
can be rewritten in the ADM formalism (up to a total derivative) as
\be
\label{ehlag}
S_{EH}=\frac{1}{16\pi\newton}\int d^4 x\,\sqrt{g}\,N\left\{
\left(K_{ij}K^{ij}- K^2\right)+R-2\Lambda\right\}.
\ee
In order to compare these two theories, it is natural to express our model in 
relativistic coordinates by rescaling $t$, 
\be
x^0=ct,
\ee
with the emergent speed of light given by
\be
c=\frac{\kappa^2\mu}{4}\sqrt{\frac{\lambdaw}{1-3\lambda}}.
\ee
Here we have assumed that $\lambdaw/(1-3\lambda)$ is positive, which is also 
required in order for the sign in front of the scalar curvature term in 
(\ref{dfpt}) to match general relativity.  Note that from the perspective 
of the $z=3$ theory at short distances, the dimension of $c$ is 
\be
[c]=2,
\ee
resulting in $[x^0]=-1$, in accord with the expected relativistic scaling in 
the infrared.  

In the rescaled coordinates $(x^0,x^i)$ suitable at long distances, the 
infrared limit of (\ref{dfpt}) then takes the general relativistic form 
(\ref{ehlag}), up to higher-derivative corrections which are suppressed at low 
energies.  The effective Newton constant is given by
\be
\newton=\frac{\kappa^2}{32\pi c},
\ee
and the effective cosmological constant
\be
\Lambda=\frac{3}{2}\lambdaw.
\ee
It is intriguing that the effective speed of light $c$, the effective Newton 
constant $\newton$ and the effective cosmological constant $\Lambda$ of the 
low-energy theory all emerge from the relevant deformations of the deeply 
nonrelativistic $z=3$ theory which dominates at short distances.  

In theories satisfying the detailed balance condition, the quantum properties 
of the $D+1$ dimensional theory are usually closely related to the quantum 
properties of the associated theory in $D$ dimensions, with action $W$.  It 
is interesting that in our case of $3+1$ dimensional gravity theory with 
detailed balance, both the Newton constant and the cosmological constant 
originate from the couplings in the action of topologically massive gravity 
in three Euclidean dimensions, a theory with excellent ultraviolet 
properties.  

\subsubsection{Soft Violations of the Detailed Balance Condition}

There is another possibility that leads to a broader spectrum of relevant 
deformations of the $z=3$ theory, without completely abandoning the 
simplifications implied by the detailed balance condition.  Starting with 
the $z=3$ theory at short distances, we can add relevant operators directly to 
the short-distance action $S$ given in (\ref{fuact}),
\be
\label{soft}
S\to S+\int dt\,d^3\bx\,\sqrt{g}\left(-M^6+\mu^4 R+\ldots\right),
\ee
with $M$ and $\mu$ arbitrary couplings of dimension 1, and ``$\ldots$'' 
denote other relevant terms with more than two spatial derivatives of the 
metric.  

This step will break the detailed balance condition, but only softly, by 
relevant operators of lower dimension than those appearing in the action at 
short distances as defined in (\ref{fuact}).  In the UV, the theory still 
satisfies detailed balance.  At long distances, the theory described by 
(\ref{soft}) again flows to $z=1$.  

\section{Other Dimensions and Values of $z$}
\label{secother}

Even though the main focus of the present paper is on the theory of gravity 
with $z=3$ in $3+1$ spacetime dimensions, the ideas are applicable in a 
broader context.  One application of the $z=2$ gravity in $2+1$ dimensions,  
as a candidate membrane worldvolume theory, was discussed in \cite{mqc}.  
Here we take at least a brief look at a list of other interesting values of 
$z$ and spacetime dimensions.

\subsection{Gravity with $z=4$ in $4+1$ Dimensions}

Power-counting renormalizability in $4+1$ dimensions requires $z=4$.  
Theories with $z=4$ satisfying the detailed balance condition in $4+1$ 
dimensions can be constructed from Euclidean gravity actions $W$ quadratic 
in curvature, familiar from the study of higher-derivative theories in $3+1$ 
dimensions.  (See, \eg , \cite{ftreport} for a review of higher-derivative 
gravity and supergravity.)  As in the case of $z=3$, we begin with first 
listing all terms of highest order in spatial derivatives, as  these are 
expected to dominate at short distances, near the hypothetical $z=4$ fixed 
point that we are attempting to construct.  The four-dimensional Euclidean 
action quadratic in curvature is given by
\be
\label{eufour}
W=\int d^4\bx\,\sqrt{g}\left(\alpha C_{ijk\ell}C^{ijk\ell}+\beta R^2\right).
\ee
This theory has two independent dimensionless couplings $\alpha$ and $\beta$.  
Modulo topological invariants, this is the most general four-derivative action 
for relativistic gravity in four dimensions.  There is no independent 
$R_{ij}R^{ij}$ term in the action, because 
\be
\int d^4\bx\,\sqrt{g}\left(R_{ijk\ell}R^{ijk\ell}-4 R_{ij}R^{ij}+R^2\right)
\ee
is a topological invariant (measuring the Euler number of the spatial slices 
$\Sigma$), as a consequence of the Gauss-Bonnet theorem in four dimensions.

We use $W$ to construct the potential $S_V$ of quantum gravity 
with $z=4$ in $4+1$ dimensions.  The high-energy limit of this theory will 
again be described by
\be
\label{genac}
S=S_K-S_V=\frac{1}{2}\int dt\,d^4\bx\,\sqrt{g}N\left\{\frac{4}{\kappa^2}
\left(K_{ij}K^{ij}
-\lambda K^2\right)-\frac{\kappa^2}{4}\frac{\delta W}{\delta g_{ij}}
\CG_{ijk\ell}\frac{\delta W}{\delta g_{k\ell}}\right\},
\ee
with $W$ now given by (\ref{eufour}).  $\kappa$ is dimensionless, as are 
the two couplings $\alpha$ and $\beta$ inherited from $W$.  This action 
can be modified by relevant operators, of dimension ${}<8$.  If we insist 
that the deformed theory satisfy detailed balance, such relevant terms in 
$S$ are generated by adding relevant operators of dimension ${}<4$ to $W$.  
Either way, the theory in $4+1$ dimensions will be dominated at long distances 
by the lowest-dimension operators in $S$, which are again given by the scalar 
curvature $R$ and the cosmological constant term.  The theory flows naturally 
to $z=1$, with an emergent speed of light, Newton constant and cosmological 
constant.   

The $z=4$ theory in $4+1$ dimensions is power-counting renormalizable.  If 
the ``quantum inheritance principle'' holds for the class of models satisfying 
the detailed balance condition described in (\ref{genac}), the 
renormalization of $\alpha$ and $\beta$ would be the same as in the 
four-dimensional relativistic higher-derivative theory described by 
(\ref{eufour}), which is believed to be asymptotically free 
\cite{stelle,ft1,ft2}.  As we mentioned in the introduction, the asymptotic 
freedom of (\ref{eufour}) would seem to make this theory an excellent 
candidate for solving the problem of quantum gravity in $3+1$ dimensions, were 
it not for one persistent flaw:  After the Wick rotation to $3+1$ dimensions, 
the spectrum of physical states contains ghosts which violate unitarity in 
perturbation theory.  

Our construction of $z=4$ theory in $4+1$ dimensions benefits from the 
asymptotic freedom of the four-dimensional higher-curvature theory 
(\ref{eufour}), but avoids the pitfall of its perturbative non-unitarity.  
Indeed, we are only interested in the four-dimensional action $W$ in the 
Euclidean signature, in order to construct the $4+1$ dimensional action 
(\ref{genac}).  

The only remaining coupling-constant renormalization in the high-energy 
limit of the theory in $4+1$ dimensions is the renormalization of $\kappa$.  
However, $\kappa$ is not an independent coupling associated with 
interactions; instead, it survives in the non-interacting limit, and 
parametrizes a family of free-field fixed point as $\alpha$ and $\beta$ are 
sent to zero.  In this respect, the quantum behavior of this theory would 
be very similar to the behavior in quantum critical Yang-Mills studied in 
\cite{cym}, which inherits asymptotic freedom from relativistic Yang-Mills 
in four dimensions. 

Setting $\beta=0$ in (\ref{eufour}) and $\lambda=1/4$ in (\ref{genac}) would 
lead to a theory which exhibits an additional gauge invariance, acting on the 
fields as
\be
g_{ij}\to\exp\left\{2\Omega(t,\bx)\right\}g_{ij},\qquad
N\to\exp\left\{4\Omega(t,\bx)\right\}N,\qquad 
N_i\to\exp\left\{2\Omega(t,\bx)\right\}N_i. 
\ee
These are the local anisotropic Weyl tranformations with $z=4$.  

\subsection{$z=4$ Gravity in $3+1$ Dimensions}
\label{secsuper}

In three dimensions, the action of Euclidean gravity quadratic in the 
curvature tensor is 
\be
\label{quadthree}
W=\frac{1}{M}\int d^3\bx\sqrt{g}\left(R_{ij}R^{ij}+\beta R^2\right).
\ee
As in four dimensions, there are again only two independent terms in $W$, but 
for a different reason:  When $D=3$, the Riemann tensor is determined 
in terms of the Ricci tensor, and the Weyl tensor vanishes identically.  
The two couplings $M$ and $M/\beta$ are now dimensionful, of dimension $1$.  
In power counting, this makes the theory described by (\ref{quadthree}) 
super-renormalizable.  When we use $W$ to generate the potential term $S_V$ 
for $z=4$ gravity in $3+1$ dimensions, we consequently end up with a 
theory whose action again has the form (\ref{genac}), now with $W$ given by 
(\ref{quadthree}) and in $3+1$ dimensions, where it is super-renormalizable by 
power counting.   As in all the previous examples with various values of $z$, 
relevant deformations flow the theory to $z=1$ in the infrared.  

Such super-renormalizable terms can also be added to our $z=3$ theory of 
gravity described in (\ref{fuact}).  These terms will give spatial dynamics 
to the conformal factor of the spatial metric, improving the short-distance 
properties of the propagator for the scalar mode $H$ of the metric, restoring 
power-counting renormalizability in the case when $H$ is present as a physical 
field.  

\subsection{The Case of $z=0$: Ultralocal Gravity}

In the Hamiltonian formulation of general relativity, the Hamiltonian is 
given by a sum of constraints,
\be
H=\int d^D\bx\left(N\CH_\perp+N^i\CH_i\right).  
\ee
Notably, the algebra of the Hamiltonian constraints $\CH_\perp(\bx)$ and  
$\CH_i(\bx)$ in general relativity is not a true Lie algebra -- in particular, 
the constraints do not form the naively expected algebra of spacetime 
diffeomorphisms.  Instead, the structure ``constants'' of the commutator of 
$\CH_\perp(\bx)$ with $\CH_\perp(\by)$ are field dependent, because they 
contain the components of the spatial metric.  

In \cite{teitelboim}, an alternative theory of gravity was proposed, in which 
the constraints do form a Lie algebra.  In this theory, the commutators of 
$\CH_i$ with themselves and with $\CH_\perp$ are the same as in general 
relativity, but the problematic field-dependent commutator of 
$\CH_\perp(\bx)$ with $\CH_\perp(\by)$ is simply replaced by zero.  This 
symmetry can be viewed either as a contraction of the symmetries formed by the 
Hamiltonian constraints of general relativity, or as a contraction of the 
algebra of infinitesimal spacetime diffeomorphisms.  The contracted symmetry 
algebra respects a dimension-one foliation of spacetime by a congruence of 
time-like curves.  This congruence can be used to identify the points of 
space at different times; as a result, the spacetime in this theory of gravity 
carries a preferred structure of absolute space.    

The theory of gravity that realizes this symmetry structure is known as the 
``ultralocal theory'' of gravity.  It is interesting to note that ultralocal 
gravity fits naturally into our framework of gravity models with anisotropic 
scaling and nontrivial dynamical exponents $z\neq 1$.  As shown in 
\cite{teitelboim}, the required symmetries force the action of the ultralocal 
theory to be of the same form $S=S_K-S_V$ as the theories considered here, 
with the potential term containing only the cosmological constant, 
\be
\label{ultra}
S_V=\int dt\,d^D\bx\,\sqrt{g}\,\Lambda,
\ee
and no curvature-dependent terms.  
There is a clear way how to interpret (\ref{ultra}) in our framework of 
gravities with anisotropic scaling:  The value of $z$ can be read off as 
one half of the number of derivatives appearing in $S_V$.  This is equivalent 
to declaring (\ref{ultra}) to be of the same dimension as the kinetic term 
$S_K$.  Either way, this approach suggests that the ultralocal theory 
corresponds formally to the limiting case of $z\to 0$.  

Historically, the ultralocal theory of gravity has been studied for at least 
two additional reasons, besides the context of \cite{teitelboim}:  
\begin{itemize}
\item[(i)] Ultralocal gravity was proposed by Isham in \cite{isham}, in an 
attempt to introduce a new formal expansion parameter into general 
relativity.  In \cite{isham}, the suggested expansion parameter was the 
coefficient in front of the scalar curvature term in $S_V$, equal to one in 
the potential of general relativity and set equal to zero in the ultralocal 
theory.  
\item[(ii)] Ultralocal theory is relevant for early universe cosmology in 
general relativity, because it captures the dynamics of FRW solutions 
in the so-called ``velocity dominated'' early stages after the big bang, as 
was first shown by Belinsky, Khalatnikhov and Lifshitz \cite{bkl1,bkl2}.  
\end{itemize}

Unfortunately, the $z\to 0$ limit is rather singular, and the program outlined 
in (i) was never very successful.  As to (ii), the embedding of ultralocal 
gravity into our framework of gravity with anisotropic scaling raises the 
possibility of interpreting the cosmological evolution as a flow, from 
$z\neq 1$ in the early universe to $z=1$ observed now.  

It is remarkable that even though the action of ultralocal theory is not 
invariant under all spacetime diffeomorphisms, the theory exhibits 
``general covariance'' \cite{teitelboim,henneaux}:  In particular, the number 
of local symmetry generators per spacetime point is $D+1$, \ie , the same as 
in general relativity.

\subsection{Bulk-Boundary Correspondence in Gravity at a Lifshitz Point}
\label{secbb}

The availability of gravity models with nontrivial values of the dynamical 
critical exponent $z$ can enhance the spectrum of examples of dualities 
between gravity in the bulk and field theory on the boundary.  This could 
be particularly relevant for understanding gravity duals of nonrelativistic 
CFTs.  

After the Wick rotation of the $z=3$ theory in $3+1$ dimensions to imaginary 
time $\tau$, the action of this theory was rewritten in a simple form 
(\ref{bact}) with the use of an auxiliary field $B^{ij}$.  The same rewriting 
applies to a much broader class of gravity models which satisfy detailed 
balance with some $D$-dimensional action $W$, such as the $z=4$ models 
discussed above.  Using this formalism, we can find a large class of classical 
solutions of such theories, simply by noting that if the following equation 
holds, 
\be
\label{bps}
\frac{1}{N}\left(\p_\tau g_{ij}-\nabla_iN_j-\nabla_jN_i\right)
-\frac{\kappa^2}{2\sqrt{g}}\CG_{ijk\ell}\frac{\delta W}{\delta g_{k\ell}}=0,
\ee
the full equations of motion are automatically satisfied.  While the full 
equations of motion are of second-order in time derivatives and of order 
$2z$ in spatial derivatives, the simpler equation (\ref{bps}) has its degree 
reduced by half.  (This argument is reminiscent of the BPS condition in 
supersymmetric theories.)  A simple class of solutions to (\ref{bps}) can 
now be obtained by setting $N=1$, $N_i=0$, and taking $g_{ij}=g_{ij}(\bx)$ to 
be an arbitrary ($\tau$-independent) solution of the equations of motion 
\be
\frac{\delta W}{\delta g_{ij}}=0
\ee
of the $D$-dimensional theory whose action is $W$.   Clearly, this 
solution can be trivially continued back to real time, and represents a 
real static solution of the full theory.

In particular, let us assume that the Euclidean action $W$ is such that it has 
the Euclidean $AdS_D$ as a solution.  This situation is rather generic, and 
does not pose a very strong restriction on $W$.  With this assumption, 
the $D+1$ dimensional theory will have a classical solution which is the 
direct product of the time dimension and $AdS_D$,
\bea
N&=&1,\quad\qquad N_i=0\vphantom{\frac{1}{2}},\cr
g_{ij}\,dx^idx^j&=&d\rho^2+\sinh^2\rho\,d\Omega_{D-1}^2.
\eea
The boundary of this solution is $S^{D-1}\times\R$.  The isometries of the 
Euclidean $AdS_D$ induce conformal symmetries in the boundary.  In addition, 
there is one more bulk isometry, given by time translations.  Thus, the 
full symmetries are
\be
\label{bulkiso}
SO(D,1)\times\R .
\ee
These symmetries suggest that the such a gravity theory in the bulk can serve 
as a possible holographic dual of dynamical field theories which are already 
critical in the static limit.  Such problems are often encountered in the 
theory of dynamical critical phenomena.  Starting with a universality class of 
a static critical system in $D-1$ spatial dimensions, the time-dependent 
dynamics of the system in $D$ dimensions can also exhibit criticality, with 
the characteristic property of ``critical slowing-down'' of time-dependent 
correlation functions.  One given static universality class can belong to 
several different dynamical universality classes.  In particular, one 
universal characteristic of the dynamics is given by the critical exponent 
$z$.  

If we study such a dynamical critical system on $\R^D$, it will exhibit 
the anisotropic scaling symmetry given by (\ref{scaling}), with $i=1,\ldots 
D-1$.  Another possibility is to put this system on $S^{D_1}\times\R$, with 
the spatial slices of the foliation given by $S^{D-1}$ of a fixed radius.  
On such a foliation, the scale symmetry (\ref{scaling}) is absent, since 
it would change the radius of the sphere.  However, the system still exhibits 
the symmetries of conformal transformations of $S^{D-1}$ and time 
translations.  Thus, the conformal symmetries left unbroken by the foliation 
are precisely the bulk isometries (\ref{bulkiso}) of the $AdS_{D}\times\R$ 
solution of gravity theory with anisotropic scaling.     

Following \cite{son,mcg}, a nonrelativistic version of the AdS/CFT 
correspondence has indeed received a lot of attention recently.  The focus 
in this area has been primarily on the CFTs with nontrivial values of $z$ 
which exhibit conventional relativistic gravity duals.  It is natural to 
broaden this framework, and free the gravity side of the duality of the 
contraints imposed by relativistic invariance.  The gravity models with 
$z\neq 1$ whose study is initiated in this paper (and in \cite{mqc}) are 
potential candidates for describing interesting universality classes on the 
CFT side, and it would seem unwise to limit the attention to CFTs with 
relativistic gravity duals.  

\section{Conclusions}

In this paper, we presented a class of gravity theories with deeply 
nonrelativistic scaling at short distances, characterized by dynamical 
critical exponent $z$.  In particular, we constructed a theory which satisfies 
the detailed balance condition with $z=3$ in $3+1$ dimensions.  This 
anisotropy between space and time improves the UV behavior of the models,  
compared to general relativity.  Moreover, such theories flow naturally at 
long distances to an effective theory with relativistic scaling and $z=1$, 
and can therefore serve as candidates for a short-distance completion of 
general relativity or its infrared modifications.  

In this picture, Lorentz invariance is only emergent at long distances, while 
the fundamental description of the theory is deeply nonrelativistic.  At short 
distances, the  spacetime manifold is equipped with an extra structure, of 
a fixed codimension-one foliation by slices of constant time.  
This preferred foliation of spacetime defines a global causal structure.  The 
existence of such a preferred causal structure puts some of the fundamental 
puzzles of general relativity and quantum gravity into a new perspective.  
In particular, various aspects of the ``problem of time'' 
\cite{ishamtime,kuchar} traditionally associated with the attempts to quantize 
general relativity are eliminated:   The preferred foliation of spacetime 
leads to an invariant notion of time, susceptible only to time-dependent 
reparametrizations.  

The existence of the preferred foliation of spacetime also changes the concept 
of black holes, and consequently the role of the information paradox and the 
holographic principle.  In general relativity, black holes are defined as 
objects with an event horizon, a notion associated with the full causal 
structure of the entire spacetime history.  Theories of gravity with 
anisotropic scaling and $z>1$ at short distances are still expected to have 
solutions that describe compact objects.  Since such theories generically flow 
at large distances to the relativistic value of $z=1$, such compact solutions 
will likely resemble the black hole solutions of general relativity (or its 
infrared modifications).  However, the notion of an event horizon for such 
solutions is emergent, and holds only approximately in the low-energy regime 
where the higher-derivative corrections to the equations of motion can be 
neglected.  At short distances, the spacetime is equipped with a preferred 
time foliation and a causal structure, which precludes the existence of event 
horizons, at least for foliations without singularities.%
\footnote{This picture might change if we allow sufficiently singular 
foliations, for example if such singularities turn out to be required for a 
consistent summation over spacetime topologies.}

If the notion of an event horizon is an emergent low-energy concept, the 
interpretation of the holographic principle also changes.  The holographic 
principle is often interpreted as a stringent bound of the number of degrees 
of freedom in a given volume of a gravitating system, as implied by the 
Bekenstein-Hawking entropy carried by black holes of the same size.  In a 
theory which is well approximated by general relativity with $z=1$ at long 
distances, but changes its scaling to $z>1$ with a preferred spacetime 
foliation at high energies, the notion of a holographic entropy bound applies 
only to degrees of freedom carrying sufficiently low energies, and should 
be viewed as an emergent feature of the low-energy dynamics.  The high-energy 
degrees of freedom can evade the bound.  

There is an intuitive way how to understand the possibility that the 
holographic bound might be an emergent low-energy bound. Recall that in the 
Bekenstein-Hawking entropy formula, the entropy of black holes is given in 
terms of the area $A$ of the horizon and the fundamental constants $c$, 
$\newton$ and $\hbar$ by
\be
S_{BH}=\frac{c^3 A}{4\newton\hbar}.
\ee
In particular, the speed of light appears in the numerator.  If the speed 
of light is effectively going to infinity at short distances (which is the 
behavior found in our anisotropic gravity models), the holographic entropy 
bound becomes less constraining at higher energies:  It only limits the number 
of possible low-energy degrees of freedom, in the regime where the behavior 
of the system is approximately relativistic.  

This behavior, leading to a radical reduction of the number of degrees of 
freedom at low energies, is very reminiscent of a similar phenomenon, 
sometimes referred to as ``rigidity,'' in ordered phases of condensed matter 
systems (see for example \cite{anderson}).  Notably, in various examples 
studied in condensed matter, this rigidity at low energies is often 
accompanied by an emergent relativistic dispersion relation for the low-energy 
excitations.  

Another aspect of gravity which might be strongly affected by the 
anisotropic scaling at short distances is cosmology.  In the high-energy 
regime relevant at early times, the effective speed of light in gravity 
models with anisotropic scaling approaches infinity, and the spacetime 
manifold exhibits the preferred foliation by constant time slices.  This 
modification of the laws of gravity changes the notion of locality and 
causality in the early stages of the universe, and can lead to new 
perspectives on the puzzles usually solved by inflationary scenarios.   

\acknowledgments
Results reported in this paper were presented at the Sowers Workshop on 
{\it What is String Theory?\/} at Virginia Tech (May 2007); at the Banff 
workshop on {\it New Dimensions in String Theory\/} (June 2008); at the 
{\it AdS, Condensed Matter and QCD\/} workshop at McGill (October 2008); and 
in talks at LBNL (December 2006), BCTP (March 2007), KITP (November 2007), 
Stanford (December 2007), Masaryk University (July 2008) and 
MIT (October 2008).  I wish to thank the organizers for their hospitality, 
and the participants for stimulating discussions.  This work has been 
supported by NSF Grant PHY-0555662, DOE Grant DE-AC03-76SF00098, and the 
Berkeley Center for Theoretical Physics.  

\bibliographystyle{JHEP}
\bibliography{lif}
\end{document}